\documentclass[sigconf]{acmart}
\usepackage{epsfig,endnotes}
\usepackage{verbatim}
\usepackage{enumitem}
\usepackage{algorithm}

\usepackage{amssymb}
\usepackage{caption}
\usepackage{subcaption}
\usepackage{mathtools}
\usepackage{algpseudocode}
\usepackage{tikz}
\usepackage{svg}
\usepackage{amsmath}
\usetikzlibrary{positioning}
\usepackage{subcaption}
\usepackage{tabularx}

\title{Ten Years of ZMap}

\begin{document}

\date{}

\author{Zakir Durumeric}
\affiliation{
    \institution{Stanford University}
    \city{Stanford}
    \state{CA}
    \country{USA}
}

\author{David Adrian}
\affiliation{
\institution{Independent Researcher}
    \city{Denver}
    \state{CO}
    \country{USA}
}

\author{Phillip Stephens}
\affiliation{
    \institution{Stanford University}
    \city{Stanford}
    \state{CA}
    \country{USA}
}

\author{Eric Wustrow}
\affiliation{%
    \institution{University of Colorado Boulder}
    \city{Boulder}
    \state{CO}
    \country{USA}
}

\author{J.~Alex~Halderman}
\affiliation{
\institution{University of Michigan}
    \city{Ann Arbor}
    \state{MI}
    \country{USA}
}

\renewcommand{\shortauthors}{Zakir Durumeric, David Adrian, Phillip Stephens, Eric Wustrow, \& J. Alex Halderman}

\thispagestyle{empty}

\begin{abstract}
Since ZMap's debut in 2013, networking and security researchers have used the open-source scanner to write hundreds of research papers that study Internet behavior.
In addition, ZMap has been adopted by the security industry to build new classes of enterprise security and compliance products.
Over the past decade, much of ZMap's behavior---ranging from its pseudorandom IP generation to its packet construction---has evolved as we have learned more about how to scan the Internet. In this work, we quantify ZMap's adoption over the ten years since its release, describe its modern behavior (and the measurements that motivated changes), and offer lessons from releasing and maintaining ZMap for future tools.
\end{abstract}




\copyrightyear{2024}
\acmYear{2024}
\setcopyright{acmlicensed}\acmConference[IMC '24]{Proceedings of the 2024 ACM Internet Measurement Conference}{November 4--6, 2024}{Madrid, Spain}
\acmBooktitle{Proceedings of the 2024 ACM Internet Measurement Conference (IMC '24), November 4--6, 2024, Madrid, Spain}
\acmDOI{10.1145/3646547.3689012}
\acmISBN{979-8-4007-0592-2/24/11}
\maketitle

\section{Introduction}

In 2013, Durumeric et al.\ released ZMap~\cite{durumeric2013zmap}, an open-source network scanner that made it dramatically easier to scan the entire IPv4 address space. Since then, more than 300~research papers have used ZMap to uncover protocol flaws~\cite{adrian2015imperfect,aviram2016drown}, shed light on the WebPKI~\cite{durumeric2013analysis}, reverse engineer mercenary spyware~\cite{marczak2014governments}, understand headline events like Heartbleed~\cite{durumeric2014matter} and Mirai~\cite{antonakakis2017understanding}, and more. Beyond research, security companies have developed products on top of ZMap to continuously monitor organizations' attack surfaces and their third-party dependencies. At the same time, like many security tools, ZMap has also been adopted by attackers to identify vulnerable systems. In aggregate, ZMap now accounts for over one-third of all Internet-wide scan traffic.

Since its initial release, ZMap has evolved as we learned more about how to scan the Internet and better understood researchers' needs. Yet, with the exception of Adrian et~al.'s write up of ZMap's 10GbE re-architecture~\cite{adrian2014zippier}, the project team has not documented these changes in the research literature. Of the improvements Adrian et~al.\ introduced, both the lock-free randomization algorithm and fast packet transmission approach have since changed. As we maintained and improved ZMap, we also learned a great deal about how to (and not to) build Internet measurement tools. 

Motivated by requests from the community to document what we have learned~\cite{claffy2019workshop}, in this work, we present a retrospective analysis of ZMap. We cover how ZMap has been adopted (\S\ref{sec:usage}),
advances in Internet scanning that challenged the tool's initial design assumptions (\S\ref{sec:lessons}), significant changes to ZMap's behavior (\S\ref{sec:changes}), and what we learned maintaining and improving ZMap (\S\ref{sec:recs}). We hope that by analyzing this endeavor, we can help the researchers who build the next generation of Internet measurement tools.

\section{Usage and Internet Landscape}
\label{sec:usage}

Since ZMap's release in 2013, its adoption has steadily increased~\cite{durumeric2014internet,anand2023aggressive}. Today, over 33\% of \emph{all} Internet-wide IPv4 scan traffic can be fingerprinted as coming from ZMap. In this section, we provide a high-level overview of ZMap's adoption by academic researchers, security companies, and malicious actors. 

\begin{figure}[h]
    \centering
    \includegraphics[width=\linewidth]{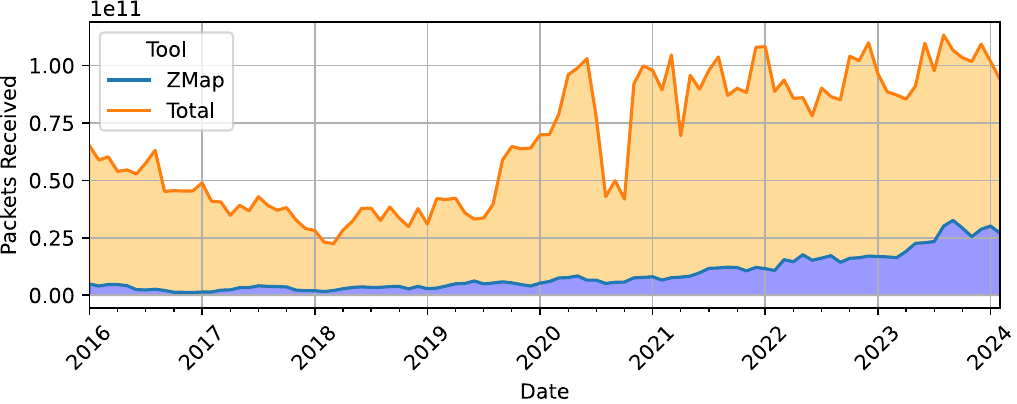}
    \vspace{-20pt}
    \caption{ZMap-Attributed TCP Scan Traffic---\textnormal{ZMap growth has accelerated significantly since 2020. In Q1 2024, 35\% of Internet-wide IPv4 TCP scan traffic (by packet) came from ZMap.}}
    \label{fig:packets_received_over_time}
    \vspace{-15pt}
\end{figure}

\subsection{Empirical Analysis}
\label{sec:darknet}

One year after ZMap's release, in 2014, Durumeric et~al.\ found relatively little adoption and that ZMap was primarily used to study academically interesting protocols like HTTP(S)~\cite{durumeric2014internet}. More recently, in 2021, Anand et~al.\ noted that many ``aggressive'' scans were from ZMap or Masscan~\cite{anand2023aggressive}. Using the same methodology as these two studies~\cite{durumeric2014internet, anand2023aggressive}, we measure ZMap adoption by analyzing scans that target at least ten IPs in the ORION Network Telescope~\cite{orion}. Our analysis is limited to TCP scans, since ORION identifies TCP, ICMP and UDP scanning flows, but only tags scanning tools for TCP flows. In addition, we note that forks of ZMap that remove the static identifying IP ID of 54321 will not be attributed to ZMap.

\begin{figure}
    \centering
    \vspace{-10pt}
    \includegraphics[width=\linewidth]{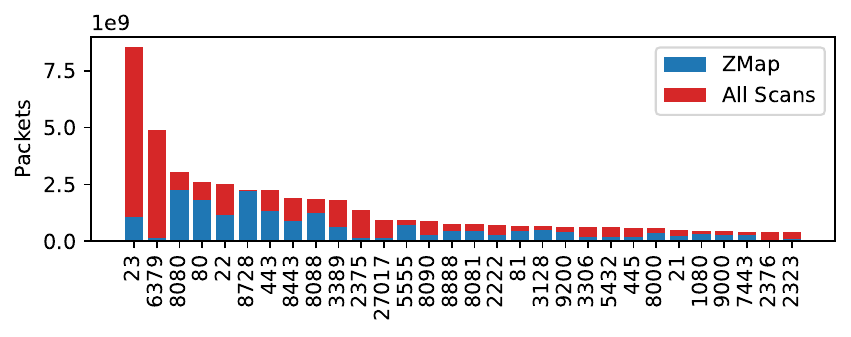}
    \vspace{-25pt}
    \caption{All TCP Scans (Top Ports by Packet)}
    \label{fig:all_ports}
    \vspace{-13pt}
\end{figure}

\begin{figure}
    \centering
    \includegraphics[width=\linewidth]{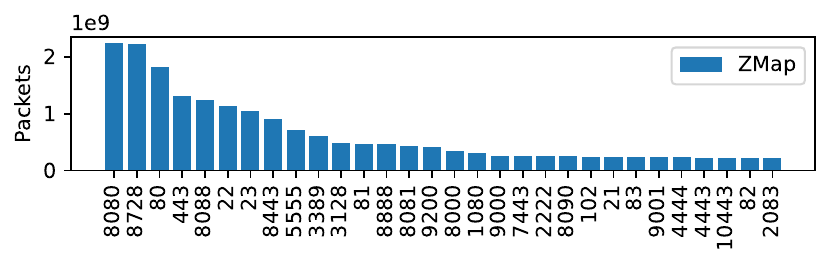}
    \vspace{-25pt}
    \caption{ZMap Scans (Top Ports By Packet)}
    \label{fig:zmap_ports}
    \vspace{-10pt}
\end{figure}

ZMap usage has increased dramatically over the past four years (Figure~\ref{fig:packets_received_over_time}). From January~1 to March~31, 2024, 35.4\% of all Internet-wide IPv4 TCP scan packets originated from ZMap. While it is possible that ICMP and UDP scans use other tools, the vast majority of scan traffic is TCP-based and ZMap still accounts for 33\% of all scan packets with only TCP traffic attributed. 
ZMap scans follow a different distribution of targeted ports than other scans (Figures~\ref{fig:all_ports} and~\ref{fig:zmap_ports}). For example, ZMap accounts for only 12\% of TCP/23 but 69\% of TCP/80 and 73\% of TCP/8080. In the most extreme case, 99.5\% of traffic targeting TCP/8728 (MikroTik router API) is from ZMap, driving it to the sixth most scanned port. There are also dramatic regional differences (Figure~\ref{fig:countries}). For example, while ZMap accounts for more than 66\% of scan packets from U.S. hosts, less than 0.5\% of Russian scan packets are from ZMap. The outsized U.S. use is driven by its adoption by American security companies (\S\ref{sec:industry}). In parallel to our work, Griffioen et~al.\ conducted an in-depth analysis of how scanning has evolved over the past decade~\cite{scanning-ten}; the study found that 59\% of Internet-wide \emph{scans} in 2024 used ZMap.


\subsection{Academic Research}
\label{sec:research}

ZMap has been used for a vast range of research purposes, from showing the possible compromise of RSA keys through transient faults~\cite{sullivan2022open} to measuring NAT64 deployment~\cite{hsu2024first}. To understand the studies that ZMap has enabled, one author manually analyzed 1,034~papers that cite or reference ZMap through April~2024 via Google Scholar and categorized papers using thematic analysis~\cite{braun2012thematic}. We exclude dissertations (since these are often comprised of published papers) as well as studies that used Censys~\cite{durumeric2015search}. In total, we identified 307~research papers directly based on ZMap data. 

While ZMap is a general measurement tool, it has most prominently been used by the security community (Appendix~\ref{app:cites}). Notably, ZMap has been used in 38~papers to uncover protocol weaknesses in TLS and underlying cryptographic primitives~\cite{beurdouche2015messy, adrian2015imperfect, aviram2016drown}, and to uncover deployment challenges and measure adoption~\cite{nemec2017measuring, kranch2015upgrading, holz2015tls}. Collecting X.509 certificates, 28~papers have shed light on the WebPKI prior to the widespread adoption of Certificate Transparency~\cite{vandersloot2016towards}. There is also a large contingent of papers that have measured the exposure of IoT devices (25~papers), ICS (14~papers), and security-relevant services (12~papers). Beyond understanding deployment patterns, a number of papers have been able to identify real-world attacks~\cite{durumeric2015neither}, to reverse engineering attacker infrastructure~\cite{antonakakis2017understanding,marczak2015pay,marczak2014governments}, and to conduct large-scale notifications using ZMap~\cite{li2016exploring, durumeric2014matter}. We encourage the security community to embrace research that builds Internet measurement tools and techniques since these are frequently used to understand and improve security.

Networking-focused papers cover topics like DNS (24~papers), BGP/RPKI (12~papers), censorship (14~papers), and IP usage and NAT (10~papers). In addition to these studies, 53~other papers reference the recommended practices when conducting measurements. 

{
\begin{figure}
\centering\small
\begin{tabular}{cccccccccc} 
\toprule
\textbf{US} & \textbf{NL} & \textbf{RU} & \textbf{DE} & \textbf{GB} & \textbf{BG} & \textbf{CN} & \textbf{IN} & \textbf{ZA} & \textbf{HK} \\
66\% & 33\% & 0.48\% & 18\% & 69\% & 9\% & 2\% & 12\% & 0.1\% & 2\% \\
\bottomrule
\end{tabular}
\vspace{-8pt}
\caption{ZMap by Country---\textnormal{The ten countries that emanate the most Internet scan traffic by packet have varied ZMap usage.
}}
\label{fig:countries}
\vspace{-10pt}
\end{figure}
}

Despite the plethora of publications, academic networks are not responsible for most ZMap traffic, likely because research experiments do not require continuous long-term scanning of a large number of ports. None of the top 100~ASes that emit the most ZMap traffic belong to universities; rather, most traffic originates from security companies and cloud providers. For example, the provider responsible for---by far---the most ZMap scan traffic is Google Cloud (GCP). Examining the reverse DNS records for scanning IPs, we find that GCP is predominately used by Palo Alto Networks to power their Xpanse Attack Surface Management product.

\subsection{Industry Adoption}
\label{sec:industry}

To understand industry adoption, we categorized the organizations identified by Greynoise as using ZMap and mapped these to broad industry categories  of security products (e.g., as defined by Gartner):

\vspace{2pt}
\noindent
\textbf{Attack Surface Management.}\quad With the shift to cloud-based infrastructure and a rise of ransomware attacks against enterprise services (e.g., MoveIt and VMWare ESXi), there has been an increased demand for companies to understand their Internet-facing infrastructure. Palo Alto Xpanse, Microsoft RiskIQ, and Rapid7 insightVM, along with numerous other smaller companies, use ZMap as the basis for providing ``attack surface management'' products that give enterprises up-to-date data about their Internet-exposed risks and potentially unknown assets. %

\vspace{2pt}
\noindent
\textbf{Third-Party Risk Management.}\quad Building on the observation that externally visible security configuration and patching patterns correlate with data breaches and compromise~\cite{liu2015cloudy, zhang2014mismanagement, liu2015predicting}, companies such as BitSight and FICO use ZMap to build security ratings that enable companies to understand their supply-chain security. %

\vspace{2pt}
\noindent
\textbf{Internet Intelligence.}\quad Several non-profits and companies use ZMap to collect data about IP addresses, threat actors, and Internet services, including BinaryEdge, Censys, IPInfo, and ShadowServer. Using this data, multiple countries proactively monitor for and notify organizations about risks (e.g., U.K.~\cite{ncsc-censys}).

\subsection{Malicious Use}
\label{sec:attackers}

Most security tools have the potential for both helping defenders and being misused by attackers. ZMap is no exception and there is evidence that ZMap has been used maliciously. 
While it is difficult to ascertain the intent of network traffic from shared providers without application-layer visibility~\cite{hiesgen2022spoki, izhikevich2023cloud}, past darknet analysis has shown that attackers have used  ``bulletproof'' hosting providers to carry out scans for vulnerable services, including MSSQL, RDP, and Mikrotik's router API~\cite{durumeric2014internet, anand2023aggressive}. Anecdotally, attackers have repurposed ZMap to carry out DOS attacks~\cite{russia-dos}, and, recently, two IoT botnets incorporated ZMap into their malware: between 2021--2023, variants of the Mirai and Medusa botnet adopted ZMap~\cite{mirai-zmap,medusa-zmap}.

\section{Lessons in Internet Scanning}
\label{sec:lessons}

Subsequent discoveries about Internet scanning have challenged some of ZMap's original design assumptions:

\vspace{2pt}
\noindent
\textbf{Port Diffusion.}\quad Despite IANA assignment of ports to L7 protocols, Bano et~al.~\cite{bano2018scanning} first noted and Izhikevich et~al.~\cite{izhikevich2021lzr} more formally showed that protocols run across a long tail of ports: only 3\% of HTTP and 6\% of TLS services run on ports~80 and~443, respectively~\cite{izhikevich2021lzr}. Scanning only assigned ports works well for understanding common user-facing protocols such as HTTP(S) but underestimates the impact of some security phenomenon, such as malware and industrial control system exposure. This has spurred new research into more intelligent Internet scanning approaches~\cite{sarabi2021smart, izhikevich2022predicting, song2023doors, luo2024ipreds} and led us to shift ZMap's address generation from being purely ``horizontal'' to support multiple ports.

\vspace{2pt}
\noindent
\textbf{L4 vs.\ L7 Discrepancies.}\quad Several studies have noted significant discrepancies between L4 and L7 responsiveness~\cite{mirian2016internet, durumeric2013analysis, holz2015tls, springall2016ftp, heninger2012mining, perino2018proxytorrent}. Izhikevich et~al.\ showed that TCP liveness does not reliably indicate service presence because of pervasive middlebox deployment~\cite{izhikevich2021lzr}. ZMap's design also encouraged what Hiesgen et~al.\ term ``two-phase scanning''~\cite{hiesgen2022spoki}, in which L4 service discovery and L7 service interrogation are performed separately. In response, some hosts ``shun'' two-phase scanners, exacerbating the perceived differences between L4 and L7 results~\cite{izhikevich2021lzr}. Sattler et~al.\ later devised a method for more accurately identifying highly L4-but-not-L7 responsive prefixes~\cite{sattler2023packed}. These differences fundamentally limit ZMap's utility (as a standalone L4 tool) to discovering \emph{potential} services, requiring most work to be completed in follow-up L7 scans and shifting our focus to downstream tools like LZR~\cite{izhikevich2021lzr} and ZGrab~\cite{zgrab2}. 

\vspace{2pt}
\noindent
\textbf{Visibility and Consistency.}\quad One challenge of running Internet scans is the lack of ground truth for validation. Wan et~al.\ showed that the ZMap paper slightly {overestimated} the visibility achieved by the tool, and that a single-probe scan actually misses about 2.7\% of HTTP(S) hosts~\cite{wan2020origin}.
Figures~1 of Hastings et~al.~\cite{hastings2016weak} and Chung et~al.~\cite{chung2016measuring} show that different organizations using ZMap sometimes see different results. However, in even the most egregious cases (e.g., Censys~\cite{durumeric2015search}), vantage points miss under 5\% of services due to blocking; the bulk of loss is typically driven by a handful of small service and cloud providers~\cite{wan2020origin}. For those who need comprehensive coverage, Wan et~al.\ recommends that the best way to mitigate transient drop is to scan from 2--3 geographically and topologically diverse vantages, rather than to send multiple probes from a single scanner, since both probes are oftentimes lost.
Results can also differ across scanning tools: Adrian et~al.\ showed that, despite following a similar high-level approach, Masscan~\cite{graham2014masscan} finds notably fewer hosts than ZMap, likely due to biases in its randomization algorithm~\cite{adrian2014zippier}.

\section{ZMap Codebase}
\label{sec:changes}

When we released ZMap, we had little idea what community, if any, would emerge. Over the past ten years, more than 80~individuals have committed code to ZMap---we are deeply thankful to those contributors. Despite the high number of committers, 90\% of ZMap code has been written by five individuals. Most contributions to the code have been made by industry and academic involvement has been limited: of the 11~external contributors who changed more than 100~LoC, only one is an academic researcher. Academic groups most frequently contributed probe modules or bugfixes; when academics made improvements to core functionality, improvements tend to be forked and renamed (e.g., XMap~\cite{xmap} and ZMapv6~\cite{gasser2016scanning}, which implemented the same IPv6 functionality) rather than upstreamed. Funding from the NSF \emph{Internet Measurement Research: Methodologies, Tools, and Infrastructure} (IMR) program~\cite{nsf-imr} has been critical in supporting ZMap's continued development. 

Beyond usability improvements and bug fixes, there have been several fundamental changes in how ZMap operates beyond the original paper's description.
We describe these changes below.

\subsection{Address and Port Generation}

One of ZMap's key contributions was its ability to \emph{statelessly} and \emph{pseudorandomly} scan the IPv4 address space. A scanner's randomization approach can dramatically affect results~\cite{adrian2014zippier} and our randomization algorithm has changed repeatedly since ZMap's initial release. ZMap originally scanned all IPv4 addresses (``horizontal scanning'') on a single port by iterating over the cyclic group $(\mathbb{Z}/(2^{32}+15)\mathbb{Z})^{\times}$, and coverting group elements to destination addresses. Soon after, we added additional smaller prime order groups to support efficiently scanning subsets of the address space (e.g., $2^{24}+43$ and $2^{16}+1$). Motivated by the findings of Izhikevich et~al.~\cite{izhikevich2021lzr}, we recently added support for multiple ports by iterating over prime order groups up to size $2^{48} + 23$ wherein the top $\lceil\text{log}_2 \mathit{IPs}\rceil$ bits of each group element are used to identify the the target IP address and bottom $\lceil\text{log}_2 \mathit{Ports}\rceil$ bits identify the target port. While conceptually approach, it has several ramifications:

\vspace{2pt}
\noindent
\textbf{Hosts vs.\ Targets.}\quad
ZMap originally tracked statistics by IP address. The new randomization approach selects from a pool of (IP, port) ``targets'', rather than considering IPs and ports independently. As a result, all data, metadata, and configuration in ZMap are now based on the notion of an (IP, port) target.
This enables randomization across the IP--Port space  (e.g., rather than rotating scans across ports with each port operating independently), but precludes options like ``max hosts'' without significant additional state. 

\vspace{2pt}
\noindent
\textbf{Identifying Generators.}\quad
To create a new permutation of the address space for every scan, ZMap originally identified a random generator (i.e., primitive root) of the appropriate multiplicative group by identifying a generator of ($\mathbb{Z}_{p-1}$, +) and then mapping it into ($\mathbb{Z}^{*}_{p}$, $\times$) ~\cite{durumeric2013zmap}. This was practical because a generator of the additive group is any integer coprime with $p-1$, which is efficiently testable with the Euclidean algorithm against randomly drawn integers. There are $\phi(2^{32}+14) \approx 10^9$ generators of the additive group, resulting in an average four attempts to identify a generator. Nearly all additive generators could be mapped into usable generators in the multiplicative group, since the only constraint on the multiplicative generators was that they were less than $2^{32}$ (to ensure safe multiplication using 64-bit arithmetic). 

For multiport scans, to support iterating over $2^{48}$ elements, we need to efficiently find multiplicative generators smaller than $2^{16}$ in a $2^{48}$ search space. Unfortunately, because an additive generator can map to a multiplicative generator anywhere in the group, only $1/2^{32}$ candidate additive generators are usable. 
To address this, we flipped our approach. For each group defined by prime $p$, we precalculate and store the prime factorization of $p-1 = k_1^{a_1} k_2^{a_2} \dots k_n^{a_n}$ for distinct primes $k_i\,(1 \leq i \leq n)$. At runtime, we generate a random candidate generator $g \in [2\mathrel{{.}\,{.}}\nobreak2^{16}-1]$ that is guaranteed to keep future arithmetic within the 64-bit address space. Then we ensure that $g$ is a generator of $\mathbb{Z}^{*}_{p}$ by checking that: $g^{(p-1)/k_i}\mod p \neq 1\,\forall\, i \in [1\mathrel{{.}\,{.}}\nobreak n]$. This requires an average four attempts.

\begin{figure}[t]
  \centering
  \includegraphics[width = 0.9\linewidth]{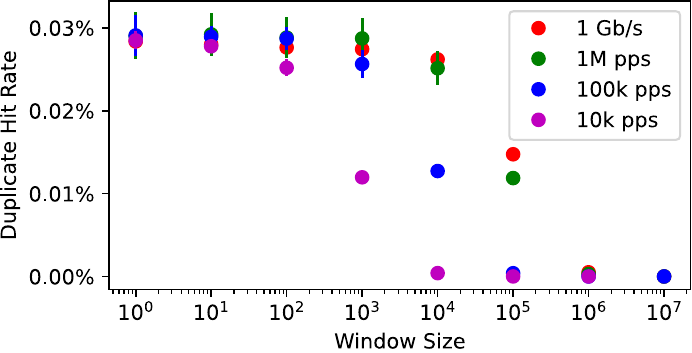}
  \vspace{-5pt}
  \caption{Sliding Window Duplicate Rate---\textnormal{We moved to a sliding window approach for deduplicating responses to support multiple ports. A window size of $10^6$ eliminates nearly all duplicates.}}
  \label{window-size}
  \vspace{-10pt}
\end{figure}

\vspace{2pt}
\noindent
\textbf{Response Deduplication.}\quad
Hosts frequently send back repeated responses, in some cases indefinitely. While we initially thought these were due to broken TCP implementations, Goldblatt et~al.\ showed that some hosts will aggressively send tens of thousands of response packets, which they term ``blowback''~\cite{goldblatt2023blowback}. We originally filtered out duplicate responses using a paged $2^{32}$-bit bitmap, which used 512\,MB of memory. While this approach guarantees no duplicates, extending it to the 48-bit space of IPs and ports would require 35\,TB. Instead, we switch to maintaining a sliding window of the last $n$ IP/Port responses, using a Judy array~\cite{baskins2000judy}. As can be seen in Figure~\ref{window-size}, a window of $10^6$~entries (ZMap default) is effective to filter nearly all duplicate responses, and lower scan rates can make do with smaller window sizes. We found zero duplicates for three trials of 1\,Gbps scans targeting TCP/80 in April~2024. 

\subsection{Scan Sharding}

In 2014, Adrian et~al.~\cite{adrian2014zippier} introduced a mutex-free sharding mechanism for ZMap's address generation. This enabled scans to be split across machines and improved performance by allowing multiple send threads on one machine to operate independently. As Mazel et~al.\ noted when they showed that ZMap can be fingerprinted through its IP generation method~\cite{mazel2019identifying}, we shifted the sharding approach in 2017. Given that other work is analyzing ZMap's address generation, we describe the two approaches:

\begin{figure}[t]
   \centering
   \begin{subfigure}{0.45\columnwidth}%
     \includegraphics[width=\linewidth]{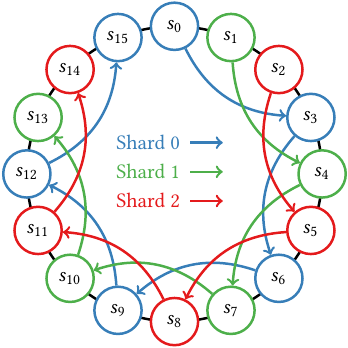}
     \caption{\textnormal{Interleaved (Old)}}
     \label{fig:sharding:interleaved}
   \end{subfigure}
   \hspace{10pt}
   \begin{subfigure}{0.45\columnwidth}
   \includegraphics[width=\linewidth]{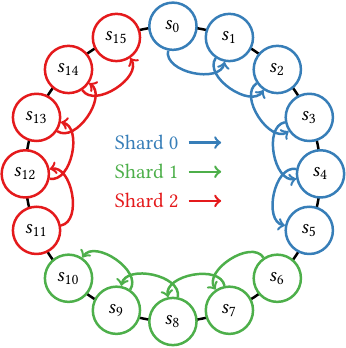}
   \caption{\textnormal{Pizza (New)}}
   \label{fig:sharding:pizza}
   \end{subfigure}
   \vspace{-5pt}
   \caption{Sharding Approaches---\textnormal{In 2017, ZMap changed sharding approaches. Replicated with permission from~\cite{mazel2019identifying}.}}
   \label{fig:zippier}
   \vspace{-10pt}
\end{figure}

\vspace{2pt}
\noindent
\textbf{Interleaved Sharding.}\quad Sharding was originally implemented with 
each shard iterating by $N$ steps at a time, offset by one step.
For $N$ shards, each shard $n \in [1\mathrel{{.}\,{.}}N]$ iterates by multiplying the current element by $g^N$, but begins iteration
at $g^n$ (Figure~\ref{fig:sharding:interleaved}).
With multiple threads in place, each shard $n$ is then further split into $T$ subshards, with each subshard
iterating by $g^{NT}$ offset by $g^{n+tN}$, where $n$ is the shard index and $t$
is the thread index. While conceptually simple, the approach requires calculating the end point of each shard to know when to stop iterating. $NT$ is not guaranteed to cleanly divide $p-1$, and so a shard might not repeat its initial value. Unfortunately, the last index of a shard does not have a closed form expression and we found that calculating it is prone to off-by-one errors. After repeated correctness issues, we switched to a simpler mechanism.

\vspace{2pt}
\noindent
\textbf{Pizza Sharding.}\quad Rather than interleaving shards, we divide the multiplicative group into $N$ ranges of values of increasing order, e.g., $[g^0,
g^{(p-1)/N})$, $[g^{(p-1)/N}$, $g^{2(p-1)/N})$, $[g^{2(p-1)/N}$, $g^{3(p-1)/N})$. For subshards, we similarly slice a single shard into $T$ ranges of values of increasing order. Visually, this is similar to slicing a pizza into $N$ slices, and then subdividing each slice into $T$ subslices (Figure~\ref{fig:sharding:pizza}). Because elements are iterated over pseudorandomly in the group, the same randomness guarantees are provided by the second approach while being easier to reason about and implement without off-by-one errors or infinite loops.

\subsection{Packet Construction}

\begin{figure}
    \centering
    \includegraphics[width=\linewidth]{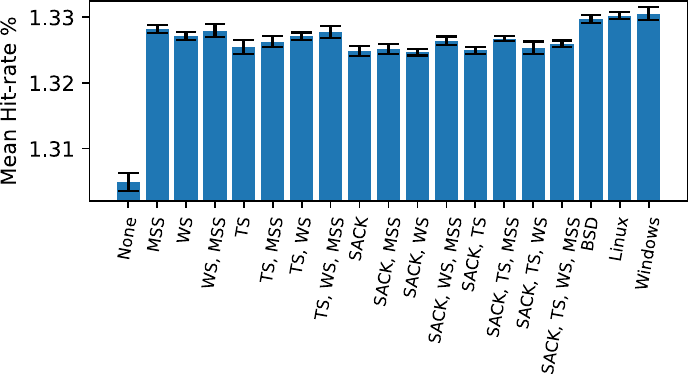}
    \vspace{-12pt}
    \caption{Hitrates for Varying TCP Options---\textnormal{SYN probes without any TCP options, as originally sent by ZMap, find 1.5--2.0\% fewer services on TCP/80 than probes that include options. Mimicking common OSes maximizes coverage. Note truncated $y$ axis.}
    }
    \label{fig:tcp-options}
    \vspace{-10pt}
\end{figure}

Striving for the highest send rate, ZMap originally used the smallest possible probes, with no included TCP or IP options. While protocol compliant, we later observed that ZMap would consistently miss some hosts accessible to OS network stacks. By varying TCP options, we found that including {any} of the Selective ACK (SA), Timestamp (TS), Window Select (WS), or Maximum Segment Size (MSS) TCP options yields a 1.5--2.0\% increase in hitrate relative to no options in a scan of TCP/80 (Figure~\ref{fig:tcp-options}). The order of options also affects results: the optimal byte-layout order, minding the TCP 4-byte word boundary, finds only slightly fewer hosts (0.0023\%, $\approx$1.5K hosts in an IPv4 scan of TCP/80) than when sending options using the exact ordering of Linux, BSD, or Windows. 

TCP options affect packet size and therefore scan rate. However, including the MSS option alone finds the vast majority of services (over 99.99\% of services on TCP/80) and remains under the minimum size of an Ethernet frame, continuing to support the maximum 1.488\,Mpps line rate of a 1\,GbE link. Using the Windows or Linux packet layouts finds slightly more services but reduces send rates to 1.389 and 1.276\,Mpps, respectively. While filtering packets based on TCP options could be due to defensive mechanisms attempting to block scanning, removing the easily identified static IP ID value of ZMap probes appears to have no impact on scan hit rates. We performed three scans of 10\% of IPv4 on TCP/80 in April 2024 with a static IP ID and with a random per-packet IP ID and find that the difference in hit-rate between the random and static IP IDs is not statistically significant. In early 2024, ZMap changed its default behavior to use random per-probe IP IDs.
\looseness=-1

\section{Lessons and Recommendations}
\label{sec:recs}

Based on our experiences, we offer several lessons and recommendations for researchers building future Internet measurement tools. These lessons are derived from decisions that we revisited, approached differently in subsequent tools like ZGrab and ZDNS, would make differently if we were to build ZMap today, or believe were fundamental to ZMap's success. As such, these recommendations are not comprehensive and are inherently opinionated, focusing on decisions where the right choice for ZMap was non-obvious to us at the time. %
They are, however, the starting point for how we would architect future measurement tools ourselves.

\vspace{2pt}
\noindent
\textbf{Tools Not Frameworks.}\quad
First documented in 1978 by McIlroy et~al.~\cite{mcilroy1978unix} and more concisely captured by Salus in 1994~\cite{salus1994quarter}, the Unix philosophy is to write programs that:

\vspace{-1pt}
\begin{enumerate}[topsep=0pt,itemsep=-0.5ex,partopsep=1ex,parsep=1ex]
    \item do one thing and do it well;
    \item work together; and 
    \item handle text streams (a universal interface).
\end{enumerate}

\noindent Nearly 50~years later, we cannot agree more with this guidance. It is difficult to predict how researchers will use measurement tools or the environments in which they will operate. ZMap was originally envisioned as a framework where researchers would build customized \emph{Scan Modules} and \emph{Output Modules} for service follow up. In practice, the vast majority of researchers use ZMap for service discovery and pipe results to secondary tools for investigation or storage. The output modules ZMap included for connecting to specific databases (e.g., Redis) became liabilities, requiring upkeep and complicating testing and packaging. In time, we removed them, opting to support only Text, CSV, and JSON Lines output.

\vspace{1pt}
\noindent \emph{Recommendation:} Build small, simple, easy-to-understand, easy-to-use, and easy-to-test measurement tools that can be creatively assembled, rather than complex applications or software frameworks. Build applications that do \emph{one} thing well. Continuously output results on a per-record/per-line basis when possible. Avoid proprietary formats and standardize output on well-worn interfaces like CSV, JSON Lines~\cite{jsonlines}, BSON, and Apache Avro. Carefully consider whether binary formats are worth the cost of direct interoperability with existing data processing toolkits and command-line tools.

\vspace{2pt}
\noindent
\textbf{Usability.}\quad
ZMap was not the first fast, asynchronous network scanner: Unicornscan~\cite{unicornscan}, Scanrand~\cite{scanrand}, and IRLscanner~\cite{leonard2010demystifying} were released years prior. IRLscanner was published at IMC; though Unicornscan and Scanrand were both unknown to the team. ZMap was likely more successful than prior tools due to its greater ease of use: it enabled researchers to scan the IPv4 Internet from a single machine by running a single command.

\vspace{1pt}
\noindent \emph{Recommendation:} Obsess over ease of installation, usage, and troubleshooting as well as documentation. It is better to have a tool that is easy to use and less full featured than vice versa.

\vspace{2pt}
\noindent
\textbf{Library and Command Line Wrapper.}\quad
It is natural to build an application where the command-line interface, application configuration, and operation are intermingled, since most tools are first used via the CLI and this is the path of least resistance. However, this design will limit a tool's potential to be integrated into larger systems. Automated or continuous measurements, such as the scans that power services like Censys~\cite{durumeric2015search}, are cumbersome to control from the CLI and more suited to a library interface.

\vspace{1pt}
\noindent \emph{Recommendation:} Structure tools with two major components: a backend library and a simple command line interface that wraps the library. This investment is relatively small and will enable the tool to be used in larger systems.

\vspace{1pt}
\noindent
\textbf{Data, Metadata, and Logs.}\quad
Given the amount of raw data collected by many measurement tools, it is difficult to tell whether experiments are operating as expected without analyzing metadata in real time. In addition, tracking as much information about the execution (e.g., time, version of software, configuration parameters, and environment) helps to later interpret, troubleshoot, or reproduce results. Logs are helpful for human debugging, but they are not inherently designed to be machine-parsable, which is needed for monitoring long-running experiments. Ultimately, we extended ZMap to produce four output streams: (1) data, (2) logs, (3) real-time updates (e.g., packets sent, received, dropped per second), and (4) machine-readable metadata at completion. 

\vspace{1pt}
\noindent \emph{Recommendation:}
Design measurement tools to produce separate streams of data, metadata, and logs. Do not cross these streams, since this complicates downstream processing. Be liberal in what environment and execution information is included in scan metadata, as it is difficult to know \emph{a priori} what will be useful. Adopt a logging library that supports multiple log levels, and use debug-level logging liberally to enable future troubleshooting. In slight contrast to SoMeta~\cite{sommers2017automatic}, we recommend that metadata collection should be built \emph{into} measurement tools to maximize ease of use.

\vspace{2pt}
\noindent
\textbf{Static Types and Output Schema.}\quad
JSON and CSV provide considerable flexibility for encoding data. For example, JSON objects can have dynamic keys and value types across records. However, downstream applications/databases often do not support this flexibility, and it is easy to create valid but painful to process records.

\vspace{1pt}
\noindent \emph{Recommendation:} Define a schema for the data you output. Ensure that each field uses a single, well defined type and that the type of one field does not depend on the value of another field. Avoid maps with dynamic keys, and instead use lists of a static document type. 
Consider using a tool like JSON Schema~\cite{json-schema} or ZSchema~\cite{zschema} to document the structure of output data and metadata.

\vspace{2pt}
\noindent
\textbf{Versioning and Releases.}\quad
We released ZMap far too infrequently, repeatedly wanting to include one more feature or bugfix in each release. In particular, ZMap~3.0 was released nearly eight years after the previous release of ZMap 2.1.1. Unfortunately, this meant that most users were either using long out-of-date releases or unversioned code. This made debugging difficult and prevented users from describing the ZMap version they used when publishing.

\vspace{1pt}
\noindent \emph{Recommendation:} Follow the Semantic Versioning Specification~\cite{semver} religiously. Focus on making regular, stable, versioned releases rather than trying to finish a preset amount of work.

\vspace{2pt}
\noindent
\textbf{Language Choice.}\quad
When we wrote ZMap, C/C++ were the only practical high-performance systems languages available. It is easy to convince oneself that it is possible to safely write C code; empirical evidence overwhelmingly says the opposite~\cite{gaynor2020science, whitehouse2024memory}. Network parsers are particularly hard to implement safely and must protect against attacker-controlled input~\cite{chromeruleoftwo}.
ZMap has had multiple regressions that caused incomplete measurements and memory safety bugs, which could have been avoided (e.g.,~\cite{bug1,bug2,bug3}). We also found that memory safety concerns make it harder to review external contributions for safety and correctness, which has reduced the rate at which we merge improvements. If were to implement ZMap today, we would do so in Rust.

\vspace{1pt}
\noindent \emph{Recommendation:} 
Develop tools in modern, memory-safe languages like Rust and Go. While Rust has a relatively steep learning curve, its safety and performance make it ideal for performance-critical applications.
Go's simple syntax and parallelism-oriented architecture make it particularly suited for quickly developing high-performance measurement tools (e.g., ZGrab~\cite{zgrab2} and ZDNS~\cite{zdns}).

\section{Internet Citizenship}
\label{sec:citizenship}

Our work provides an opportunity to revisit our original best practices for conducting scans~\cite{durumeric2013zmap}. Overall, we believe that the 2013 recommendations remain a sound set of considerations. However, we encourage the measurement community to treat these recommendations as good practices for \emph{most} research, not as a set of requirements nor as the basis for having conducted research ethically. For example, there may be situations when scans are better performed unattributed or when opting out specific networks could invalidate results (e.g., tracking a specific threat actor). We additionally recommend that researchers:%

\vspace{-4pt}
\begin{enumerate}[topsep=2pt,itemsep=-0.6ex,partopsep=1ex,parsep=1ex]

    \item Investigate whether existing datasets suffice. Often, these provide better coverage and reduce aggregate bandwidth.

    \item Publish newly collected datasets for other researchers. 

    \item Deploy WHOIS entries that identify how to contact you. 

    \item Validate how handshakes will appear in logs. For example, some benign SSH handshakes inadvertently show up as failed authentication attempts that concern operators.

\end{enumerate}

\vspace{-4pt}
\noindent Institutions have adopted different practices for validating opt-out requests~\cite{cant-stop-the-scan}. We have found it necessary to verify the authenticity of exclusion requests. Given that IP address ownership changes over time, it likely makes sense to eventually expire opt-out requests and it may not make sense for institutions to share blocklists. Our team expires requests after 1--2~years and found that in the {vast majority} of exclusions are not re-requested. We offer the following, updated set of best practices as a \emph{recommended starting point} when conducting active measurements:

\vspace{-4pt}
\begin{enumerate}[topsep=2pt,itemsep=-0.5ex,partopsep=1ex,parsep=1ex]

    \item \textbf{Minimize Internet Impact.} While Internet scanning is a powerful research methodology, it can also affect systems and create work for operators. Consider whether existing open source datasets provide the data you need. If you do perform scans, conduct scans no larger or more frequent than necessary and at the minimum scan rate needed for your research objectives. Publish any scan data you collect.

    \item \textbf{Signal Intent.} When possible, publish reverse DNS entries, IP WHOIS records, and a website that describes the scans. Ensure that operators can easily contact the research team. 
    
    \item \textbf{Provide An Opt-Out Mechanism.} Provide a simple mechanism for operators to request exclusion from future scans. Indicate the IP ranges you use for scanning so that operators can drop research traffic themselves.

    \item \textbf{Proactively Investigate Effects.} Run newly developed scanning code against your own systems to ensure that you understand how scans might affect devices and appear in logs. Start with small experiments before completing full scans in case your scanner causes unexpected problems.

    \item \textbf{Coordinate Locally.} Coordinate with your local IT and security teams to reduce the risk of overwhelming local networks, as well as to ensure that they know how to handle any inbound inquiries from operators.

    \item \textbf{Disclose Results.} When appropriate, consider how you can improve the security of the systems you have scanned. Responsibly disclose security problems you uncover and consider notifying vulnerable system owners.

\end{enumerate}
\vspace{-8pt}

\vspace{-1pt}
\section{Conclusion}

The most exciting aspect of building ZMap has been watching how other researchers have used it in unpredicted but valuable ways to meaningfully improve the Internet. We are sincerely thankful to everyone who has contributed, pushing the tool close to feature completion. While the development of new ZMap features has slowed, we are excited to continue to expand the ecosystem of tools that work with ZMap (e.g., ZDNS~\cite{zdns} and ZGrab~\cite{zgrab2}) and to enable an even broader range of measurement uses. Many of the lessons we learned from maintaining ZMap may seem obvious today, but were not obvious at the time. We hope that our retrospective analysis will help the community build an even richer and more reliable ecosystem of Internet measurement tools moving forward.

\begin{acks}
We thank all of the contributors to ZMap, the IT and Security teams at the University of Michigan and Stanford University, Michael Bailey, Jeff Cody, and Liz Izhikevich. We thank the reviewers at IMC for their suggestions and feedback. This work was supported by the National Science Foundation under grants CNS-2223360, CNS-1518888, and CNS-2319080, as well as a Sloan Research Fellowship. Any opinions, findings, conclusions, or recommendations expressed in this material are those of the authors and do not necessarily reflect the views of their employers or the sponsors.
\end{acks}

{\footnotesize \bibliographystyle{acm} \balance
\bibliography{sample}}

\clearpage 

\clearpage

\appendix

\clearpage

\section{Ethics}
Our work is primarily a metareview of prior studies and a presentation of lessons learned from ZMap. When conducting our own experiments to validate changes made to ZMap, we followed the original guidelines set forth by Durumeric et~al.\ in 2013~\cite{durumeric2013zmap}. We provide updated recommendations on how to best conduct Internet scanning in Section~\ref{sec:citizenship}.

\section{Academic Usage of ZMap}
\label{app:cites}

\begin{figure}[h!]
    \centering
    \small
    \begin{tabular}{lrl}
    \toprule
    Topic & Papers & Examples \\
    \midrule
    Censorship and Anonymity & 14 & \cite{ramesh2020decentralized, pearce2017augur, pearce2017global, frolov2020detecting} \\
    Cryptography and Key Generation & 17 & \cite{bos2014elliptic, janovsky2020biased, vsvenda2016million, hastings2016weak} \\
    Denial of Service (DoS) & 15 & \cite{giotsas2017inferring, bushart2018dns, bock2021weaponizing} \\
    DNS and Naming & 24 & \cite{liu2016all, lu2019end, zhu2015measuring} \\
    Email and Spam & 8 & \cite{szurdi2017email, maroofi2021adoption, durumeric2015neither, holz2015tls} \\
    Exposure, Hygiene, and Patching & 12 & \cite{bonkoski2013illuminating, fiebig2016one, zhang2014mismanagement, durumeric2014matter} \\
    Honeypots, Telescopes, and Attacks & 9 & \cite{franzen2022looking, alt2014uncovering, morishita2019detect} \\
    IP Usage, DHCP Churn, Nand AT & 10 & \cite{moura2015dynamic, lee2016identifying, hsu2024first} \\
    Industrial Control Systems (ICS) & 14 & \cite{mirian2016internet, dahlmanns2020easing, dahlmanns2022missed} \\
    Internet of Things (IoT) &  25 & \cite{mangino2020internet, antonakakis2017understanding, costin2014large, srinivasa2021open} \\
    Systems and Network Security & 19 & \cite{xu2015measurement, vissers2015maneuvering, guo2018abusing, malhotra2015attacking} \\
    PKI, Certificates, Revocation & 28 & \cite{liu2015end, brubaker2014using, durumeric2013analysis, perl2014you} \\
    Power Outages and Grid Monitoring & 4 & \cite{bayat2021down, andersonpowerping} \\
    Privacy & 5 & \cite{perta2014exploiting, van2022saving, hilts2015half} \\
    QUIC & 7 & \cite{ruth2018first, zirngibl2021s, kosek2022dns} \\
    Routing, BGP, and RPKI & 12 & \cite{giotsas2015mapping, alaraj2023global, rodday2021deployment, hlavacek2020disco} \\
    Scanning and Device Identification & 25 & \cite{sattler2023packed, yang2019towards,albakour2021third, albakour2023pushing} \\
    TLS, HTTPS, and SSH & 38 & \cite{adrian2015imperfect, brinkmann2021alpaca, holz2015tls, springall2016measuring} \\
    Understanding Threat Actors & 4 & \cite{marczak2014governments, marczak2015pay} \\
    \midrule
    Other Internet Measurement Topics & 26 & \cite{mehani2015early, rye2019sundials, guo2018detecting, lichtblau2017detection} \\
    Ethics Guidance Only (No ZMap Use) & 53 & \cite{nguyen2022freely, kim2018measuring, izhikevich2024democratizing, moon2021accurately} \\
    \bottomrule
    \end{tabular}
    \vspace{2pt}
    \caption{Academic Papers Built on ZMap Data---\textnormal{We manually investigated papers that cited ZMap or ZMap-derived datasets to understand what types of research studies have used ZMap data.}}
\end{figure}

\end{document}